\documentclass[useAMS,usenatbib]{mn2e}

\include{epsf}

\usepackage{graphicx}

\title[QSO Lensing Magnification: A Comparison of 2QZ and SDSS Results]{QSO Lensing Magnification: A Comparison of 2QZ and SDSS Results}
\author[G. Mountrichas \& T. Shanks]{G. Mountrichas\footnotemark[0]\thanks{E-mail:georgios.mountrichas@durham.ac.uk} \& T. Shanks \\
Department of Physics, University of Durham, South Road, Durham DH1 3LE, UK}

\begin{document}

\date{21 March 2006}

\pagerange{\pageref{firstpage}--\pageref{lastpage}} \pubyear{2002}

\maketitle

\label{firstpage}

\begin{abstract}

The lensing magnification of background QSOs by foreground galaxies and
clusters is a powerful probe of the mass density of the Universe and the
power spectrum of mass clustering. However, the observational results in
this area have been controversial. In particular, the 2dF QSO survey
suggests that a strong anticorrelation effect at $g<21$ is seen for both
galaxies and clusters which implied that galaxies are anti-biased
($b\approx0.1$) on small scales at a higher level than predicted by the standard
cosmology (Myers et al., 2003, 2005) whereas results from SDSS suggested
that the effect was much smaller ($b\approx1$) and in line with
standard expectations (Scranton et al., 2005).

We first cross-correlate the SDSS photo-z, $g<21$, $1.0<z_p<2.2$ QSOs
with $g<21$ galaxies and clusters in the same areas. The anti-correlation found is somewhat less than the results of Myers et al. based on 2QZ QSOs. But contamination of the QSOs by low redshift
NELGs and QSOs can cause underestimation of the anticorrelation lensing
signal. Correcting for such low redshift contamination at the levels
indicated by our spectroscopic checks suggests that the effect is
generally small for QSO cross-correlations with $g<21$ galaxies
but may be an issue for fainter galaxy samples. Thus when this
correction is applied to the photo-z QSO sample of Scranton et al. the
anti-correlation increases and the agreement with the 2QZ results of
Myers et al. is improved. When we also take into account the fainter
$r<21$ galaxy limit of Scranton et al. as opposed to $g<21$ for Myers et
al., the two observational results appear to be in very good agreement.
This therefore leaves open the question of why the theoretical
interpretations are so different for these analyses. We note that the
results of Guimaraes, Myers \& Shanks based on mock catalogues from the
$\Lambda$CDM Hubble Volume strongly suggest that QSO lensing at the
levels detected by both Myers et al. and now Scranton et al. is
incompatible with a galaxy bias of $b\approx1$ in the standard
cosmological model. If the QSO lensing results are correct then the
consequences for cosmology may be significant (see Shanks 2006).

\end{abstract}


\section{Introduction}

Large concentrations of mass at low redshift such as galaxies and groups
of galaxies can gravitationally lens background objects such as
galaxies, QSOs, supernovae and the cosmic microwave background. This
phenomenon affects these background sources in two ways. It magnifies
and distorts them. This systematic distortion is called cosmic shear (see e.g. Mellier and Meylan 2005 for a review) and the magnification cosmic magnification (Wu 1994, Broadhurst et al. 1995). In turn cosmic magnification has also two effects named the solid angle and amplification effect. The amplification effect brightens the apparent magnitude of the background sources resulting in an increase of
the objects that we observe in a magnitude limited survey whereas for
the solid angle effect, the lensing effectively reduces the solid
angle behind the lens, decreasing the number of background sources.
These two competing effects are responsible for the different results
expected for different magnitude QSO samples. So for bright QSO samples
with a steep number count slope the amplification effect dominates and
we expect a positive cross-correlation with foreground galaxies whereas
for faint samples with a flatter count slope we expect a negative
cross-correlation signal. Intermediate QSO samples give a null result. For
example, Boyle et al. (1988) found significant anti-correlation on scales of 4$'$ around galaxies using faint QSOs ($B<20.9$).
Williams and Irwin (1998) and Nollenberg and Williams (2005) found significant positive correlation on angular scales of the order of one degree (their QSO samples respectively consisted of QSOs within $16<m_B<18.5$ and $13<B<17.5$) and Gaztanaga (2003) measured the cross-correlation between photometric galaxies and bright, $i<18.8$ spectroscopic QSOs using only the SDSS EDR and found a positive cross-correlation of 20\% on arcminute scales.

Although the idea seems simple, QSO-galaxy cross-correlations have been
a controversial subject over the years as different results and
different strength of the signal are detected even when the same or similar
QSO magnitude samples have been used. In this paper we are looking for a
possible explanation of the apparent discrepancy between the results of
Scranton et al. (2005) and Myers et al. (2003, 2005) where in the latter
papers they seem to find a much stronger anti-correlation signal at the
same ($g<21$) QSO magnitude limit. Our first aim will be to reanalyze
the SDSS photo-z data and compare with previous results. In particular,
we shall look for the effects of low redshift contamination in the
photo-z QSO sample and also measure the QSO-galaxy cross-correlation at
the same QSO and galaxy limits as used by Myers et al. (2003, 2005). We shall also be looking at the important effect of the $g<21$ galaxy magnitude limit 
used by Myers et al (2003, 2005) as opposed to the $r<21$ limit used by 
Scranton et al (2005).

In Section 2 we explain the data that we use and how they differ from
Scranton et al. and Myers et al. (2005) data sets, as well as providing
details of our analysis. In Section 3 we present our results from
QSO-galaxy cross-correlation and in Section 4 the results of correlating
the same QSOs with galaxy groups. In Section 5 we check the low-z
galaxy/QSO contamination in Richards et al. QSO sample that we (and
Scranton et al.) use and the effect of this contamination on the
results. Note that we use the term contamination in different contexts to include
both `catastrophic redshift failures' and also non-QSOs in the
$1<z_p<2.2$ QSO sample. In Section 6 we use fitting models for the
cross-correlations with clusters and galaxies from Myers et al. (2003
and 2005) and Scranton et al and compare them with our results. Finally,
in Section 7 we discuss the conclusions that can be drawn as to the
reason that causes the difference in the results in the three published
papers.

\begin{figure}
\begin{center}
\centerline{\epsfxsize = 9.0cm
\epsfbox{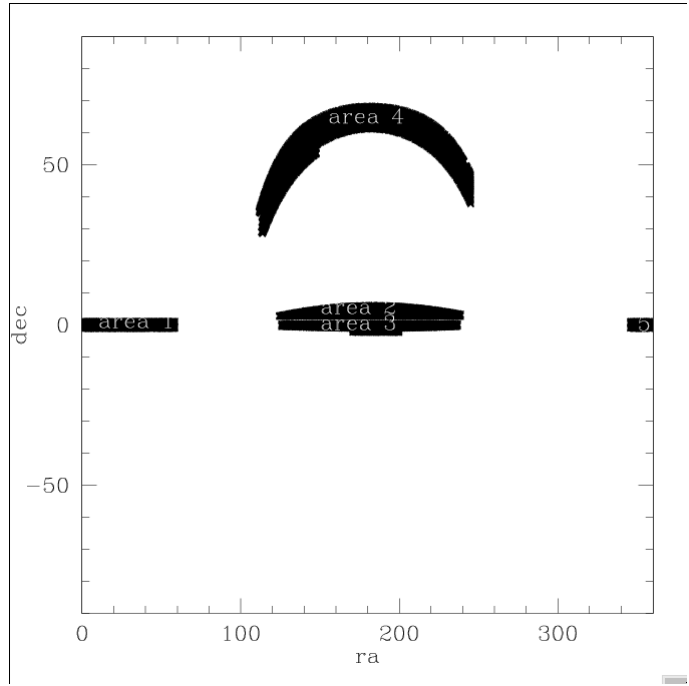}}
\caption{The distribution of our QSO and galaxy samples. The numbers indicate the areas in which the samples were cross-correlated. Area 3 is the 2QZ area.}
\label{fig:xi_s2qz}
\end{center}
\end{figure}

\begin{table}
\caption{Number of $g<21$ QSOs, $g<21$ galaxies and the QSO and galaxy density for each area separately}
\centering
\setlength{\tabcolsep}{0.5mm}
\begin{tabular}{lccccc}
       \hline
$area$ & $1$ & $2$ &  $3$ & $4$ & $5$\\
       \hline \hline
QSOs & $5,373$  & $10,380$ & $5,720$ & $15,225$ & $1,178$\\
       \hline
Galaxies & $203,164$  & $455,541$ & $216,204$ & $679,200$ & $45,050$\\
       \hline
QSOs deg$^{-2}$  & $34.5$  & $31.6$ & $31.8$ & $29.2$ & $30.2$  \\
       \hline
Galaxies deg$^{-2}$ & $1302$  & $1386$ & $1202$ & $1303$ & $1155$  \\
       \hline 
\end{tabular}
\end{table}

\section{The Data and Analysis}

Our galaxy sample consists of SDSS DR4 galaxies at the magnitude limit
of $g\leq 21$ where the sky density is $\simeq1200deg^{-2}$. It should
be noted at the outset that our galaxy sample and that of Myers et al.
(2005) is different from that of Scranton et al., as they use galaxies
with $r<21$ with a sky density of $\simeq3500deg^{-2}$. This means that
care must be taken in comparing these results because the projected
galaxy clustering in the $r<21$ sample will have $\approx2\times$ lower
correlation function amplitude due to its increased depth. The QSO
sample we use is the same sample which is extracted by Richards et al.
(2004), by applying their `Kernel Density Estimation' method
on the DR1 dataset. Scranton et al. use a similar method to extract
their QSO sample but they apply it to the DR3 set instead. We also use
the same redshift range as Scranton et al. which is $1.0\leq z_p\leq 
2.2$. So the main difference between the QSO sample we use for our
analysis and Scranton et al. use for theirs should be a larger number of
QSOs in their sample (DR3 vs. DR1).

In total we have 37,876 QSOs in the above redshift range and 1,599,159
galaxies. The numbers for each area separately as well as the galaxy
density ($g<21$) are shown in Table 1. Our random catalogue consists of
7 times the number of the galaxies. All `holes' found in SDSS data have
been added to our random catalogues so the numbers shown above are our
final number of objects. Fig. 1 shows the distribution of the QSOs that
comprise our sample. The indicated areas 1-5 are those that were used
for our cross-correlation analysis. In terms of our analyses of the SDSS
photo-z QSO samples, we accept that these may be less sophisticated
than those of Scranton et al. For example, we do not mask out poor
seeing or high reddening areas and we use the standard SDSS star-galaxy
classifier rather than Bayesian star-galaxy separation parameters. We
believe that our $g<21$ QSO limit is conservative enough to make these
differences cause negligible effects. More importantly, we ignore the
statistical ranges allowed for QSO photo-z, $z_p$, generally taking a sharp cut with
$1<z_p<2.2$. Scranton et al also weight their cross-correlation by the
QSO photo-z probability. We shall flag the points where these different 
approaches may affect our conclusions.

The 2QZ QSOs are taken from the 2QZ catalogue (Croom et al. 2004). The
2QZ comprises two $5deg\times75deg$ declination strips, one in an
equatorial region in the North Galactic Cap (NGC) and one at the South
Galactic Pole (SGC). In our analysis when we mention the 2QZ area or 2QZ
QSOs we mean only the NGC of the 2dF QSO survey, unless we present
results from Myers et al. (2005) in which case the cross-correlation has
been done in both the NGC and SGC. We should underline here that the
most important difference between the 2QZ and the DR1 QSO dataset that
we use or the DR3 that Scranton et al. use is that 2QZ QSOs are
spectroscopically confirmed whereas the method used to extract the
DR1/DR3 QSOs is based on Bayesian photometric classification and the
redshifts assigned to the objects are photometric redshifts.

Throughout our analysis we centre on QSOs and count galaxies or clusters
for our cross-correlations so our random catalogues are constructed with
the same angular selection function as our galaxy samples. To measure
the two-point correlation function $\omega(\theta)$, we use the
expression (Peebles 1980), 

\begin{equation}
\omega(\theta)=
\frac{DD_{12}(\theta)\overline{n}}{DR_{12}(\theta)}-1 
\end {equation}

where $DD_{12}$ is the number of data-data point pairs (e.g. QSO-galaxy
pairs) and $DR_{12}$ is the data-random point pairs. $\overline{n}$ is a
factor which shows how many more random points than data points we have
(7 in our case). Our errors are field-to-field errors (Myers et al.,
2003). In this case we divide our data sets into 25 subsamples and
measure the field-to-field variations of the cross-correlation function.
The error is then estimated as the standard error inverse weighted by
variance to account for different numbers of objects in each subsample,
i.e.

\begin{equation}
\sigma_\omega^2(\theta) =
\frac{1}{N-1}\sum_{L=1}^{N}\frac{DR_L(\theta)}{DR(\theta)}[\omega_L(\theta)-\omega(\theta)]^2
\end {equation}

Since 2QZ QSOs had higher priority than 2dFGRS galaxies for
spectroscopic observations there is no issue for fibre incompleteness to
deal with. Of course, this is also true for the photo-z sample. On the
smallest scales there is a lower limit for cross-correlation of 10$''$
due to confusion caused by galaxy overlaps.

\begin{figure}
\begin{center}
\centerline{\epsfxsize = 9.0cm
\epsfbox{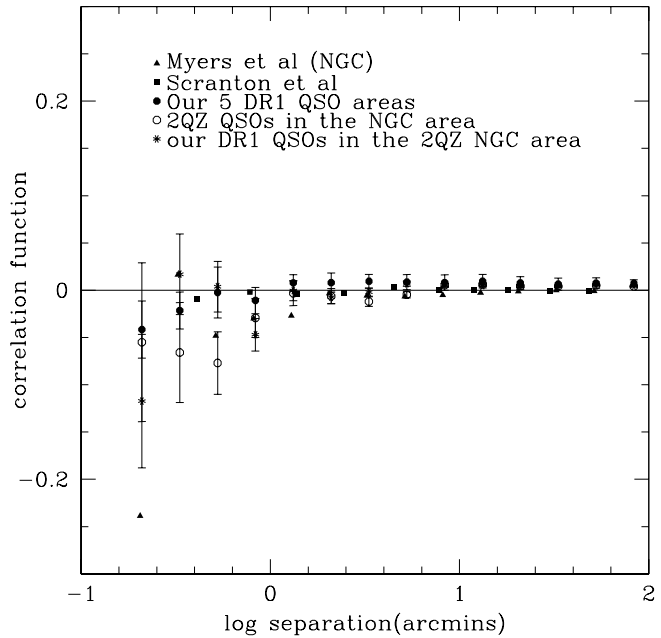}}
\caption{QSO-galaxy cross-correlation results. Our DR1 results are shown
by the black filled circles and cover the whole magnitude range for the
QSO sample ($g<21$) and $g<21$ for the galaxies. Triangles are the results from Myers et al. (2005) for the NGC of the 2QZ (centring on QSOs and counting galaxies). Squares show Scranton et al. where the faintest QSO sample has been used, $20.5<g<21$, and $r<21$ for galaxies. We have also included our results by cross-correlating 2QZ QSOs with the same galaxy sample in the 2QZ area. These results are shown by the open circles. Asterisks show the results from cross-correlating our DR1 QSOs with the same galaxy sample in the 2QZ area.
The errors are field-to-field errors.}
\label{fig:xi_s2qz}
\end{center}
\end{figure}

\begin{figure}
\begin{center}
\centerline{\epsfxsize = 9.0cm
\epsfbox{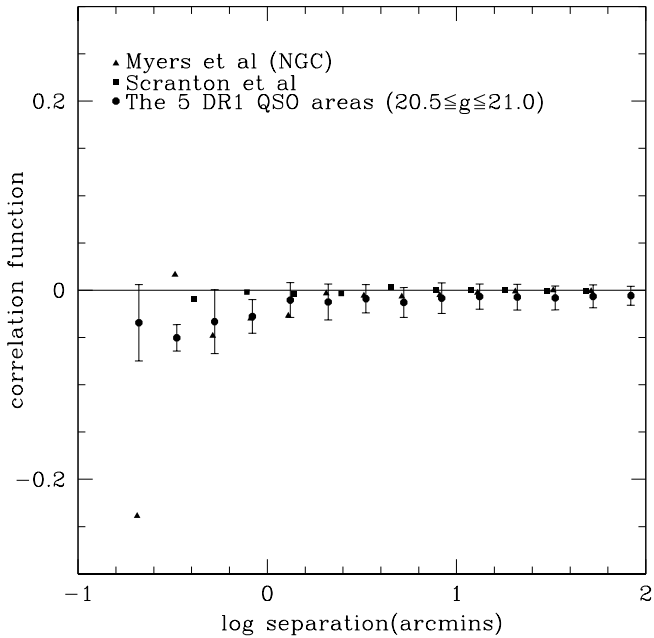}}
\caption{QSO-galaxy cross-correlation results. This time our DR1 QSO
sample comprises of QSOs with $20.5\leq g\leq 21.0$. The errors are
field-to-field errors. Triangles are the results from Myers et al. (2005)
and squares from Scranton et al.}
\label{fig:xi_s2qz}
\end{center}
\end{figure}

\begin{figure}
\begin{center}
\centerline{\epsfxsize = 9.0cm
\epsfbox{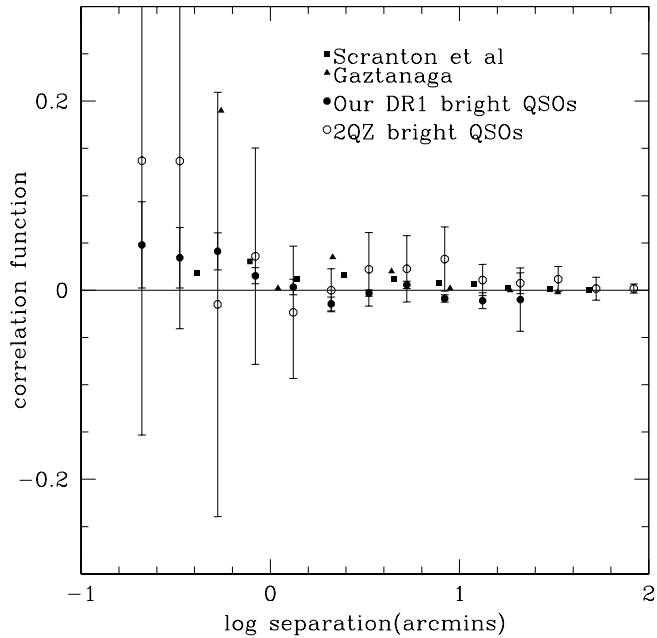}}
\caption{QSO-galaxy cross-correlation results for bright QSOs. Our DR1
QSO sample consists of QSOs with $17\leq g \leq 19$ (black circles). The
results of Scranton et al. for QSOs with $17\leq g\leq  19$ and galaxies with $r<21$ are also shown (squares). The triangles show
the results from Gaztanaga (2004). His sample consists of QSOs with
$18.3\leq i \leq 18.8$ and $0.8\leq z \leq 2.5$ and galaxies with
$19<r<22$. Finally, we present the results from the 339 bright
($18.25\leq  b_j \leq  19.0)$ 2QZ QSOs in the 2QZ area (open circles).}
\label{fig:xi_s2qz}
\end{center}
\end{figure}

\section{QSO-galaxy cross-correlation results}

We first cross-correlate our SDSS DR1 photo-z $g<21$ QSO sample with our
SDSS DR4 $g<21$ galaxy sample. The results are shown in Fig. 2 by the
black filled circles. For comparison the NGC results from Myers et al.
(2005) have also been added (triangles) without their errorbars and for
reference the results from Scranton et al. (squares) also without their
errorbars. The results from Scranton et al. are taken from their faintest
QSO sample $20.5<g<21$, cross-correlated with $r<21$ galaxies. From this
plot it seems that our results are slightly higher than those of Myers
et al. (2005) throughout most of the range of scales. The Scranton et al.
results have a smaller anti-correlation signal as expected due to the
different galaxy sample they use (Section 6).

We have also included the results when we cross-correlate 2QZ QSOs with
the same galaxy sample $(g<21)$ in the 2QZ NGC area. These results are
shown by the open circles. This is 1.8 $\times$ bigger area than used by
Myers et al. (2003, 2005) due to the smaller SDSS area previously
available. Finally, asterisks show the results from cross-correlating
our DR1 QSOs with the same galaxy sample in the 2QZ area. The 2QZ QSOs
again tend to give more anti-correlation than the DR1 QSOs
cross-correlated with the same galaxy sample.

Fig. 3 shows the same results as the previous plot but this time our DR1
QSO sample consists of 17,426 QSOs with magnitude $20.5\leq g\leq21.0$. The
amplitude of the anticorrelation signal is similar to that found in Fig.
2 for $g<21$ QSOs and again is stronger, as expected, than the signal
detected by Scranton et al. for $r<21$ galaxies, but consistent with
Myers et al.

Fig. 4 shows the QSO-galaxy cross-correlation results for bright QSOs.
Here our DR1 sample consists of $1.0<z_p<2.2$ QSOs in the magnitude
range $17\leq g\leq  19$, Scranton et al. (squares) where the QSOs have
$17\leq g\leq  19$ and the galaxies have $r<21$. We have also plotted
the results from Gaztanaga (2004) (triangles) where the specific data
points are drawn from his QSO sample with $18.3\leq i\leq  18.8$ and
$0.8\leq  z\leq  2.5$ and their galaxy sample with $19<r<22$. We then
show the results when we cross-correlate the 339 bright ($18.25\leq
g\leq  19.0$) 2QZ QSOs in the 2QZ NGC area (open circles). These results
give zero signal and a bump appears on scales from 4$'$-16$'$. This bump
is consistent with the statistical noise as we can see from the
field-to-field errorbars and moreover it disappears when we
cross-correlate the same QSO sample with galaxies in groups (Section 4,
Fig. 9). Our DR1 and 2QZ results seem consistent although the errors are
large. Both results also appear lower than the results of Gaztanaga. The
results are consistent with those of Scranton et al. although their
galaxy sample is fainter. Summarising, our results for bright QSOs, from
DR1 and 2QZ give, at least at small scales, a less positive
cross-correlation than that seen by Gaztanaga but one consistent with
that found by Scranton et al., although the S/N is poor. The
interpretation of this latter result requires account to be taken of
their fainter limit and the QSO count slope at $g\simeq19$. We
postpone further discussion to Sections 5 and 6 where the effects of
contamination will also be discussed.

\begin{figure*} 
\begin{center}
\centerline{\epsfxsize = 12.0cm 
\epsfbox{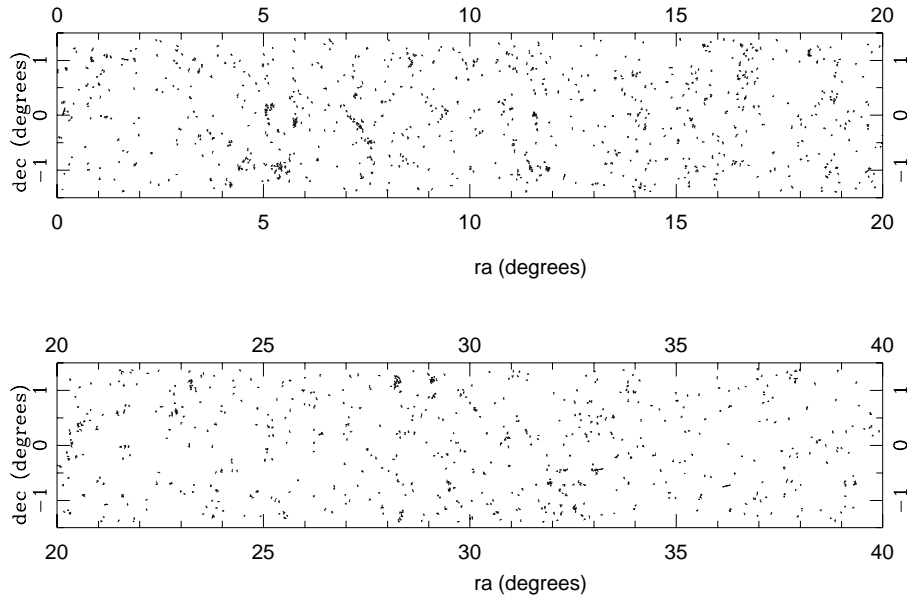}}
\caption{Galaxies ($g<21$) in groups with more than 7 members in the strip between $-1.5<\delta<+1.5$deg and 0h$<\alpha<$2h40 in Area 1. }
\label{fig:xi_s2qz}
\end{center}
\end{figure*}

\section{QSO-galaxies in groups cross-correlation results}

The next step was to find groups of galaxies in the 5 areas of the DR4
dataset. Although there are no group results from Scranton et al. with
which to compare, we can still compare the QSO photo-z and spectroscopic
results, in a situation where the S/N of any lensing effect may be
expected to be higher than for galaxies. We therefore use the same
method that is described in Myers et al. (2003) and references therein to
determine these groups. We use a factor $\delta$ by which we wish our
group density to exceed the mean surface density of the area that is
being examined. In our case we use $\delta=8$. Then we draw a circle
with the largest possible radius, such that our group density doesn't
fall below $\delta$ times the mean surface density. Groups are defined
when these circles overlap and the mean surface density does not fall
below the critical value (friends-of-friends). From these groups we
select for our cross-correlations the ones with at least 7 members in
order to reduce the likelihood of chance alignments of galaxies at
different redshifts being grouped together. An example of how the
galaxies of these groups look is shown in Fig. 5. In total there are
14,143 groups with more than 7 members in the 5 areas that we use.

We next cross-correlated our DR1 QSO sample with galaxies that are in 
groups with at least 7 members. The signal detected in this case is
usually stronger than the one in QSO-galaxy cross-correlations because
groups have a bigger mass. In total our galaxy sample consists of
146,490 galaxies in groups. In Fig. 6 we show the results for each of
the 5 areas separately. The errors are field-field errors. An
anti-correlation effect is consistently seen in all 5 areas. The combined
results for all the 5 areas are shown in Fig. 7. For comparison, the
results from Myers et al. (2003) for both the NGC and SGC have been
added (triangles). They use 22,417 2QZ QSOs and nearly 300,000 galaxies
of limiting magnitude $b=20.5$ found in groups of at least 7 members.
From the comparison we see that their signal is stronger on scales of
0.4-2.5 arcmins, which is expected as the results from the SGC show a
slightly stronger anti-correlation signal (Myers et al., 2003) and so
they push the overall result down. Again, as for the QSO-galaxy results
we show the results of the cross-correlation of 2QZ QSOs with the
galaxies in clusters in the NGC of the 2QZ area. The results are shown
by the open circles. Finally, asterisks show the results from our DR1
QSO cross-correlation with the same galaxy sample in the 2QZ area. The
signal detected is at a lower level. From the plot we see that 2QZ QSOs
give a stronger anti-correlation signal than the one detected by using
our DR1 QSOs in the 2QZ area but statistically consistent with Myers et
al (2003).

Previously we have cross-correlated individual group galaxies rather
than the centre of groups. Fig. 8 shows the results from DR1 QSOs with
centres of groups of galaxies with more than 7 members (filled circles).
Open circles show the results from Myers et al. (2003) combined for both
the NGC and the SGC together with a best fit model which will be briefly
discussed in Section 7. Although these results are generally easier to 
compare directly with models, the signal-noise is weaker due to
non-weighting by cluster membership. Again our DR1 results appear to
show less anti-correlation than the 2QZ results.

Fig. 9 also shows the results from cross-correlating our bright DR1 QSOs
($17.0\leq g \leq 19.0$) with galaxies in clusters with more than 7
members (filled circles). Some positive signal at 1$'$ is again seen, 
increased by about a factor of four from the QSO-galaxy case in Fig. 4.
However, the positive signal is still only at a marginally significant
level. We shall argue later that the lack of a strong positive
signal may be explained by being close to the knee of the QSO count slope
where $\beta\approx0.4$ and only a small lensing effect might be
expected.

Finally, we extracted the bright QSOs from the 2QZ sample ($18.25\leq
b_j \leq 19.0$, $1.0\leq z\leq 2.2$) and cross-correlate them with
galaxies in clusters (with more than 7 members). The results are shown
by the open circles in Fig. 9 and seem to give a slightly positive
signal but this result is again not statistically significant. Also, the
bump that appeared in Fig. 4 when we cross-correlated the same QSO
sample with galaxies has disappeared.

\begin{figure}
\begin{center}
\centerline{\epsfxsize = 9.0cm
\epsfbox{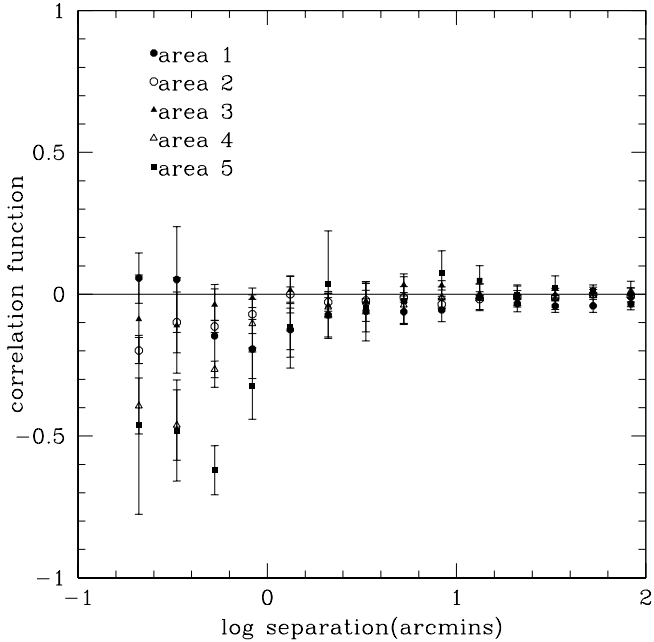}}
\caption{The DR1 QSO - DR4 galaxies in groups cross-correlation results for each area separately. The errors are field-field errors.}
\label{fig:xi_s2qz}
\end{center}
\end{figure}

\begin{figure}
\begin{center}
\centerline{\epsfxsize = 9.0cm
\epsfbox{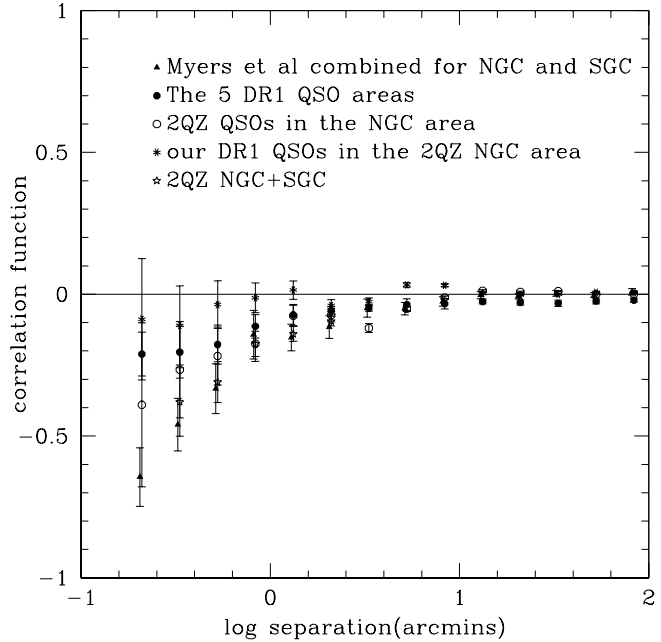}}
\caption{Cross-correlation between QSOs and galaxies in groups of
galaxies with at least 7 members (filled circles). The errors are
field-to-field errors. The triangles show the combined results from
Myers et al. (2003) for both the NGC and SGC. They use 22,417 2QZ QSOs
and nearly 300,000 galaxies of limiting magnitude $b=20.5$ found in
groups of at least 7 members. Their cross-correlation is done in the SGC
and NGC strip of the 2QZ. From the comparison we see that their signal
is stronger on scales of 0.4-2.5 arcmins. We also show the results of
the cross-correlation of 2QZ QSOs with the galaxies in clusters in the
2QZ area. The results are shown by the open circles and give a weaker
anti-correlation signal than found in the results of Myers et al. but
still the signal is stronger than the one detected by using our DR1 QSO
sample. Asterisks show the results from our DR1 QSOs cross-correlation
with the same galaxy sample in the 2QZ area. Stars show the results when
we average our open blue circles from the NGC of 2QZ with Myers et al.
results for the SGC.}
\label{fig:xi_s2qz}
\end{center}
\end{figure}

\begin{figure}
\centerline{\includegraphics[angle=0.0,scale=0.4]{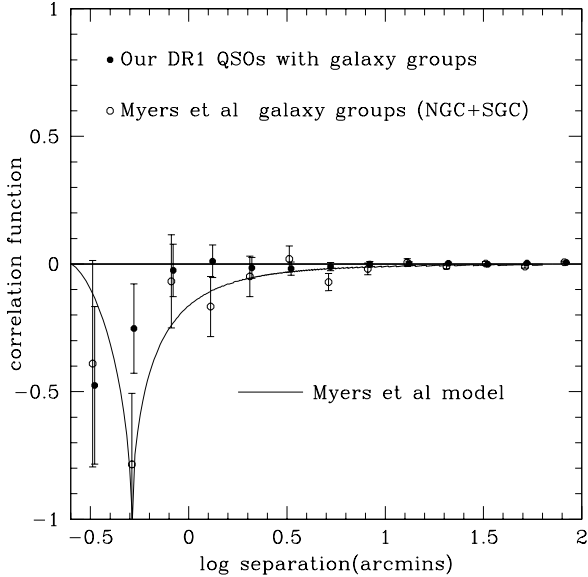}} 
\caption{Cross-correlation between DR1 QSOs in 5 areas and centres of
groups of galaxies with at least 7 members, assuming $\delta=8$. Open
circles show the results from Myers et al. (2003) combined for both the
NGC and the SGC together with a best fit model which will be briefly
discussed in Section 8. Our signal here is less strong due to non-weighting
by cluster membership. }
\end{figure}

\begin{figure}
\begin{center}
\centerline{\epsfxsize = 9.0cm
\epsfbox{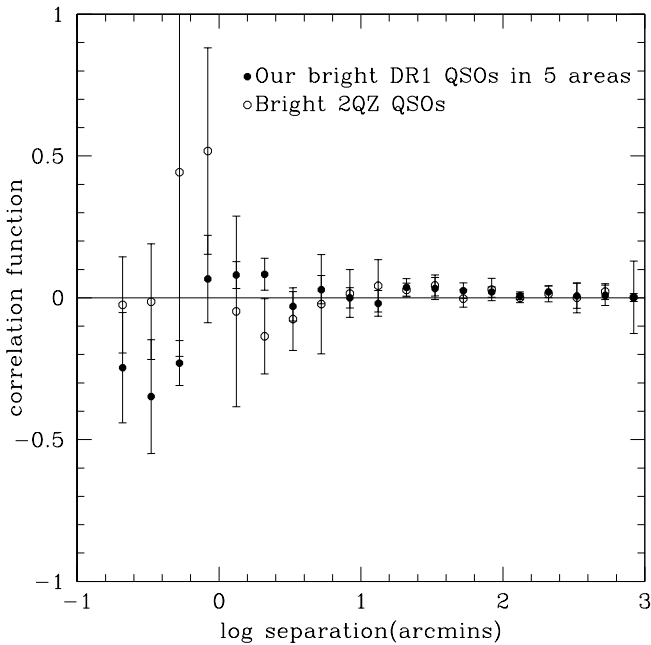}}
\caption{Bright 2QZ QSOs ($18.25\leq b_j \leq 19.0$, $1.0\leq z\leq
2.2$) cross-correlated with galaxies in clusters with at least 7 members
(open circles) give a slightly stronger positive signal but still
insignificant. Also, the bump has disappeared which suggests that it was
due to statistics. Filled circles show the results for our bright DR1
QSOs ($17.0\leq g \leq 19.0$) cross-correlated with galaxies in clusters
with at least 7 members in the 5 areas. A signal is marginally
detected at 1$'$.}
\label{fig:xi_s2qz}
\end{center}
\end{figure}

\section{Contamination in the QSO sample}

As previously noted our DR1 QSO sample consists of photometric QSOs that
are derived using the method described by Richards et al. and is similar
to the method that has been used by Scranton et al. The main difference
with the Scranton et al. sample is that their sample is extracted from
DR3 whereas the sample we use is extracted from DR1. In this Section we
examine the contamination percentage in our photometric sample and how
this contamination could change the results. The contamination that is
most important is the fraction of low redshift QSOs and compact narrow
emission-line galaxies (NELGs) in the $1.0<z_p<2.2$ QSO photo-z sample
since they can confuse any lensing signal with intrinsic clustering with
the galaxies but we shall also be include stellar contamination which
can dilute to a lesser extent the lensing signal. Richards et al. (2004)
and also Myers et al (2007) claim 5\% non-QSO contamination over the
full magnitude and redshift range. We now wish to check this number
and, more importantly, also determine the contamination of the
$1<z_p<2.2$ QSO photo-z sample by $z<0.6$ ($z<0.3$) QSOs and NELGs that
may affect cross-correlation with $r<21$ ($g<21$) galaxies.

\subsection{Comparison with the 2QZ catalogue}

In order to check the contamination in the DR1 QSO sample we first
compare it with the 2QZ catalogue. In the 2QZ there are 23,290 objects
and in our DR1 QSO sample 4,535 QSOs in the 2QZ area, with $g<20.85$,
$1.0<z<2.2$. 3,025 objects are common in the two sets. 2,516 of these
objects have been identified as QSOs in the 2QZ and 509 have different
or no ID in the 2QZ, ie 16 are NELGs, 91 stars and 402 have not been
identified by the 2QZ team. Finally, 1,510 ($=4,535-3,025$) DR1 QSOs are
not included in the 2QZ. The contamination in this case refers to
objects that have been identified as QSOs in our DR1 photometric sample
but have different IDs in the 2QZ catalogue and QSOs that have been
assigned a high photometric redshift but their spectroscopic redshift in
the 2QZ data set is lower. Fig. 10 shows the photometric vs. the
spectroscopic redshift of the common objects (including stars, NELGS and
objects with no ID). Table 2 summarises the contamination statistics as
derived from the plot. There are 169 low spectroscopic redshift QSOs and
NELGs in a total of 2024 objects. According to these numbers we find a
contamination of $(8.3\pm0.6)\%$ for objects that have photometric
redshift between $1.0<z_p<2.2$ and 2QZ spectroscopic redshift $z_s<1.0$.
In the same way the contamination of spectroscopic redshift $z_s<0.6$
objects is $(3.6\pm0.4)$\% (73 low spectroscopic redshift QSOs and NELGs
in a total of 2,024 objects). Finally, for $z<0.3$ the contamination is
$(0.6\pm0.2)\%$ (12 spectroscopic redshift QSOs and NELGs at low-z in a
total of 2,024 objects).

DR1 objects that have been identified as stars in the 2QZ also dilute
any anti-correlation by the factor $(1-f_s)^2$ where $f_s$ is the
fraction of stars in the the DR1 sample. This effect of including
uncorrelated stars is usually much smaller than the effect of including
low redshift NELGs and QSOs. From Table 2 $f_s=55/2024=$2.7\% giving
$(1-f_s)^2=0.95$ which implies that the anti-correlation is also
decreased by $\approx5$\% due to star contamination.

\begin{figure}
\begin{center}
\centerline{\epsfxsize = 9.0cm
\epsfbox{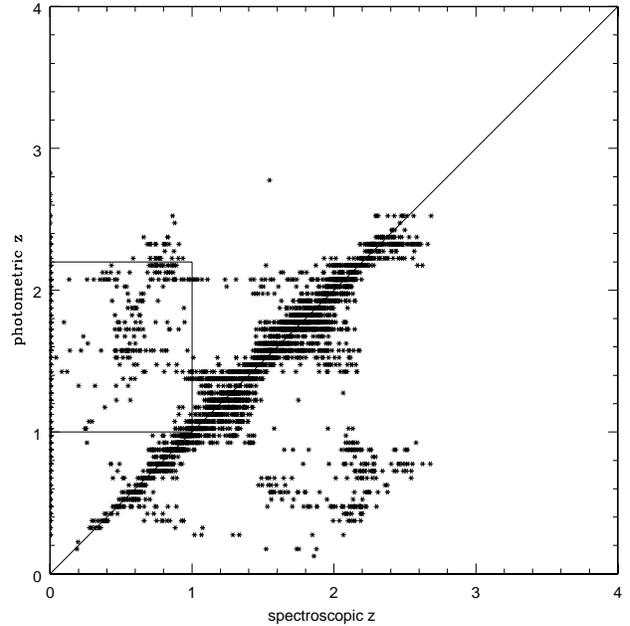}}
\caption{Photometric vs. spectroscopic redshift for the common objects
between the DR1 and 2QZ catalogues.}
\label{fig:xi_s2qz}
\end{center}
\end{figure}

\subsection{Comparison with the 2QZ+SDSS catalogue}
The DR1 QSO catalogue we are using has also spectroscopic redshifts for the
QSOs wherever available, either from the 2QZ or from spectra obtained by
SDSS. Based on this information there are 1,215 NELGs and QSOs with
photometric redshift $1.0<z<2.2$ and spectroscopic redshift $z<1.0$ out
of 15,776 objects. So the overall contamination is $(7.7\pm0.2)$\%. For
spectroscopic redshift $z<0.6$ there are 594 contaminating
QSOs and NELGs so the contamination is now $(3.8\pm0.2)$\% and finally
for $z<0.3$ there are 134 low spectroscopic QSOs and NELGs and the
contamination is $(0.9\pm0.1)$\%. At $20.5<g<21$ 39/936 are $z<0.6$ QSOs and NELGs and so the contamination here is $4.2\pm0.7$\%

\subsection{Comparison with QSOs in the COSMOS field}

Next we repeated the same procedure using the QSO spectroscopic sample
from the Prescott et al. (2006). In that paper they confirm 95 quasars from the
SDSS DR1 catalog in the COSMOS field. The quasars are within
$18.3<g<22.5$ and a range in redshift $0.2<z<2.3$. 42 out of these 95
QSOs are in our DR1 sample at its $g<21$ limit and 31 have
$1.0<z_p<2.2$. So in a similar way we plot their spectroscopic redshift
(from Prescott et al.) against their photometric redshift (from our DR1
sample) in Fig. 11. As we see there are three contaminants (open
circles) out of the 31 objects that have photometric redshift between
$1<z_p<2.2$ (asterisks), so the contamination now is $(9.7\pm5.6)$\%
with spectroscopic redshift $z<1.0$. Only one of these 3 has $z<0.6$ and
this has $z=0.0375$ which makes the contamination $(3.2\pm3.2)$\% both
for spectroscopic redshift ranges $z<0.6$ and $z<0.3$ but the sample
may be too small to draw any strong conclusion.

\begin{figure}
\begin{center}
\centerline{\epsfxsize = 9.0cm
\epsfbox{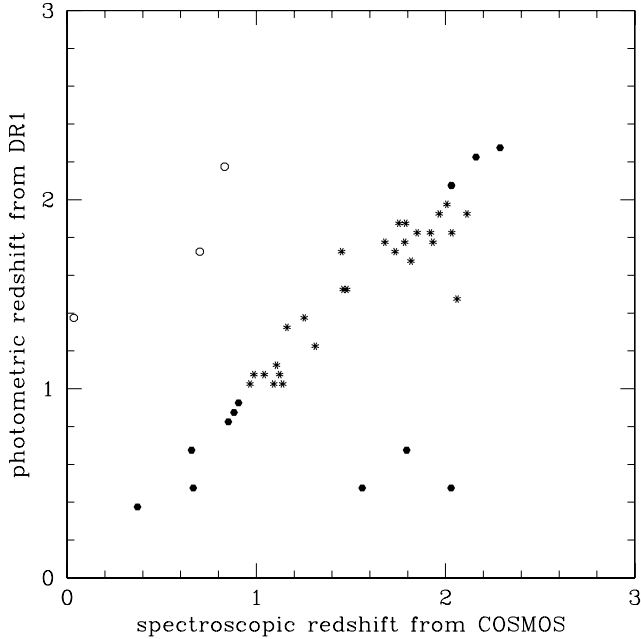}}
\caption{Spectroscopic redshifts (from Prescott et al.) against
photometric redshifts (from DR1 sample). The open circles are the three
contaminants and the asterisks are the objects with photometric redshift
$1.0<z<2.2$.}
\label{fig:xi_s2qz}
\end{center}
\end{figure}

\subsection{Comparison with spectroscopic QSOs from AA$\Omega$.}

In order to check the contamination statistics particularly of the
$\simeq\frac{1}{3}$ DR1 QSOs that lie outside the 2QZ selection at $g<21$
we made new $AA\Omega$ (Sharp et al. 2006) observations in four fields.
A full description of the observations will be given by Mountrichas et
al (2007) so they are described only briefly here. The $AA\Omega$ fields
are the COSMOS field with centre 10 00 28.8  02 12 21, the COMBO-17 S11
field with 11 42 58.0  -01 42 50, the 2SLAQ d05 field with 13 21 36.0  -00 12
35 and the 2SLAQ e04 field 14 47 36.0  -00 12 35. There are 123 QSO spectra
taken for DR1 QSOs that are not previously included in 2QZ nor in
COSMOS. In the data reduction we used the AUTOZ and 2DFEMLINES routines
written by Lance Miller and Scott Croom for the 2QZ. We first used AUTOZ
which automatically identifies objects and assigns redshifts to them.
Then, using the 2DFEMLINES program we checked these redshifts by eye.
This program steps through each object in the file created by the AUTOZ
in turn starting with the best IDs through the worst. If the ID and the
redshift are ok we moved to the next object. If either the
identification or the redshift were not secure a `?' quality flag
was put next to the ID or the redshift. From these 123 objects
identified as QSOs in our DR1 QSO photometric sample there were 5 NELGs
with $z\leq 0.3$ and 1 QSO with $z<0.6$. So from the comparison of our
photometric sample with this spectroscopic one, the contamination is
$(4.1\pm1.8)$\% for $z\leq 0.3$ and $(4.9\pm2.0)$\% for $z\leq 0.6$. For
these calculations we have only taken into account first-class spectra
that have unambiguous redshifts. That means that whenever we had to add
a question mark to the redshift then this object was excluded as a
potential contaminant but was still counted in the 123 objects. If we
exclude these objects as well from the total number of objects, then
the total number of good spectra falls to 78 and the contamination goes
up to $(6.4\pm2.9)$\% for $z\leq 0.3$ and $(7.7\pm3.4)$\% for $z\leq
0.5$. The former, more conservative, numbers are listed in Table 3.

\subsection{Summary of the contamination results} 

The contamination results summarised in Table 3 allow comparison between
the three data sets described above. From that we can conclude that for
spectroscopic redshift $z<0.3$ our DR1 sample has an $\approx2$\%
contamination. This comes from the fact that $1/3$ of our sample which
is in 2QZ has contamination $0.6$\% and the rest has $4.1\pm1.8$\% as
found from the comparison with the four AA$\Omega$ fields. Weighting by
the relative size of the 2QZ and non-2QZ components of the DR1
$1<z_p<2.2$ QSO sample, this gives an estimate of $1.8\pm0.6$\% for the
$z<0.3$ contamination. The error here is dominated by the AA$\Omega$
estimate of the non-2QZ DR1 contamination. However, we note that the
2QZ$+$SDSS contamination is higher ($0.9\pm0.1$\%) than the 2QZ
contamination ($0.6\pm0.2$\%) and that the COSMOS estimate of overall
contamination is also higher so we believe our estimate of $1.8\pm0.6\%$
for $z<0.3$ contamination is reasonable. The DR1 $z<0.6$ contamination,
which is more appropriate for $r<21$ selected galaxies, is similarly
estimated to be $(2/3\times3.6+1/3\times4.9)=4.0\pm0.7$\% (see Table 3).
The overall contamination for $z_s<1$ is estimated to be $(2/3\times8.3+ 1/3\times10.6=9.1\pm0.6$\%). This can be compared to the
low-redshift contamination rate of 7.3\% (A.D. Myers , priv. comm)
indicated by Fig. 2 of Myers et al. (2007). These contamination fractions
may only apply approximately to the QSO-photo-z sample of Scranton et al
since a more detailed comparison would need to take account of their
use of the statistical redshift range allowed by the photo-z estimates.

\subsection{Effects of $\simeq2\%$ contamination on the DR1 QSO-galaxy results}

Now we shall see how an $\approx2$\% contamination for $z\leq 0.3$ can
alter our cross-correlation results in the case of QSO-galaxy and
QSO-cluster cross-correlations. We will first base our calculations on
the galaxy-galaxy autocorrelation results at $g\leq21.0$ represented by
$w=0.33\theta^{-0.8}$ with $\theta$ in arcmins (see Fig. 8 of Myers et
al). We multiplied this $g<21$ galaxy $w(\theta)$ by the $z<0.3$ 1.8\%
contamination estimate, and then subtracted this correction from the
SDSS photo-z galaxy-QSO cross-correlation results. At 1$'$ this
correction is -0.007 and at the smallest bin $\theta=0'.2$ the
correction only rises to -0.02.

Assuming this correction, our previous Fig. 2 now appears as seen in
Fig. 12. In this plot our DR1 QSO results in the 5 areas (filled
circles) have been changed assuming 1.8\% contamination and remain
statistically consistent with the 2QZ QSO results (open circles). The
agreement improves but only slightly. Also, the results using our DR1
QSO sample in the 2QZ area (asterisks) have been changed using the same
corrections and they remain statistically consistent with the 2QZ QSO
and our DR1 QSO results.

Finally, we have corrected the Scranton et al. results (squares)
assuming contamination of 4\% and the fit to the $r<21$ galaxy
autocorrelation function shown in Fig. 17. At 1$'$ this correction is
-0.006 and at the smallest bin $\theta=0'.3$ the correction rises to
-0.01. The anti-correlation at $\theta=0'.3$ increased by a factor of
$\simeq2$.


\begin{figure}
\begin{center}
\centerline{\epsfxsize = 9.0cm
\epsfbox{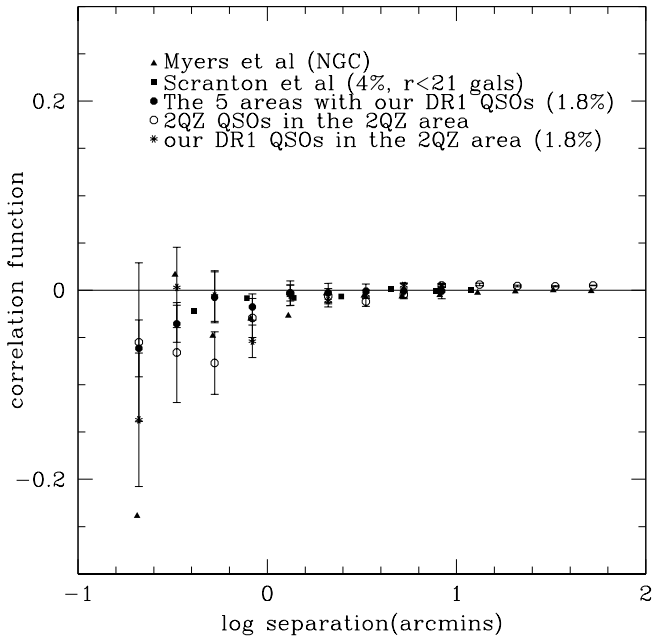}}
\caption{QSO-galaxy cross-correlation results as in Fig. 2 but our DR1
QSOs in the 5 areas (filled circles) and in the 2QZ area
(asterisks) have been changed assuming contamination of 1.8\% and the fit to $g<21$ galaxy autocorrelation function shown in Fig.17. Scranton et al. (squares) have also been changed assuming contamination of 4\% and the fit to the $r<21$ galaxy autocorrelation function (Fig.17).}
\label{fig:xi_s2qz}
\end{center}
\end{figure}

We have also corrected the results for the bright photo-z QSOs samples
that appear in Fig. 4. The corrected results are shown in Fig. 13 
assuming 1.8\% contamination. These results show some positive signal but only at marginally
significant levels. The most significant result is from the $17<g<19$ QSO-galaxy result of Scranton et al. 
which after correction gives $w_{qg}=0.024\pm0.018$ at $1'$.


\begin{figure}
\begin{center}
\centerline{\epsfxsize = 9.0cm
\epsfbox{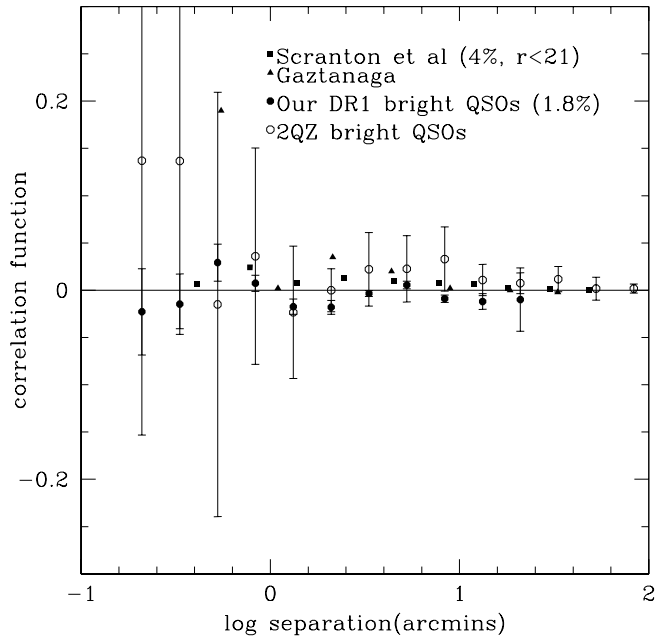}}
\caption{QSO-galaxy cross-correlation results for bright QSOs as in Fig.
4. Our DR1 QSO sample consists of QSOs with $17\leq g \leq 19$ (filled
circles) and is corrected assuming 1.8\% contamination of the $g<21$
galaxies, Scranton et al. (squares) is corrected assuming 4\%
contamination and the fit to the $r<21$ galaxy autocorrelation function
shown in Fig. 17. The triangles show the results from Gaztanaga (2004).
His sample consists of QSOs with $18.3\leq i \leq 18.8$ and $0.8\leq z
\leq 2.5$. Finally, we present the results from the 339 bright
($18.25\leq  b_j \leq  19.0)$ 2QZ QSOs in the 2QZ area (open circles).}
\label{fig:xi_s2qz}
\end{center}
\end{figure}

We now correct the QSO-cluster cross-correlations for contamination.
Here we base the correction on the galaxy-cluster cross-correlation results of Stevenson et al. (1988) who used the same group detection
parameters as we do here. They cross-correlate galaxies with groups of
galaxies with $>7$ and more than $>15$ members down to $b_j=20.2$ (see
their Fig. 9). We use their results for clusters with more than 7
members which matches our cluster selection and we correct our results
assuming as before 1.8\% contamination of galaxies in our DR1 QSO
sample. At 1$'$ this correction is -0.11 and at the smallest bin $\theta=0'.3$ the correction rises to -0.12. The results are shown in Fig. 14. Applying the 1.8\% contamination correction to our results increases the anti-correlation
at 1$'$ by a factor $\simeq3$ and improves consistency with the Myers et al.
(2003) model at all scales.

Summarising, the effects of contamination are low for the photo-z QSOs when cross-correlated with $g<21$ galaxy samples. They can be more significant at small scales for photo-z QSOs cross-correlated with $r<21$ galaxy samples.

\begin{figure}
\begin{center}
\centerline{\epsfxsize = 9.0cm
\epsfbox{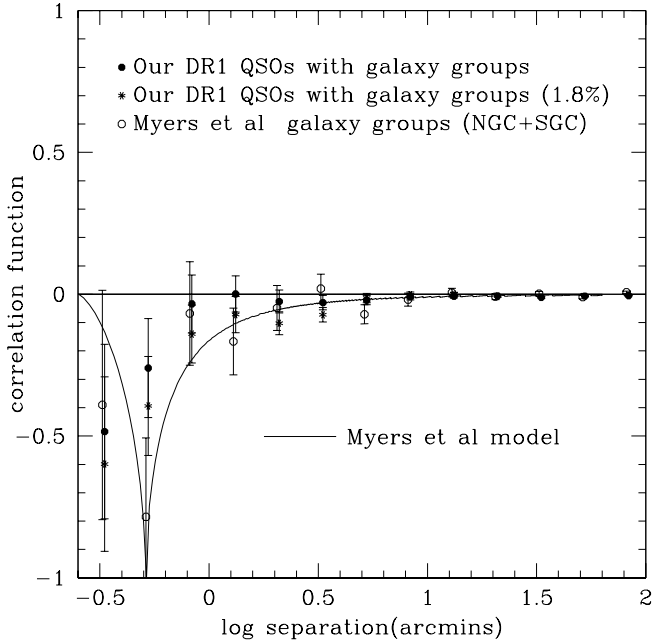}}
\caption{QSO-galaxy group centres cross-correlations. The filled
circles show the QSO-group centres cross-correlation results as shown in
Fig. 9. The asterisks show the results when we consider 1.8\%
contamination and take into account the galaxy-cluster results from
Stevenson et al. Open circles show the results of Myers et al. (2003).
The model from Myers et al. is also shown (solid line)}
\label{fig:xi_s2qz}
\end{center}
\end{figure}

\subsection{Contamination correction estimated via low-$z$ objects in the 2QZ
catalogue} 
A different way to correct for the contamination in the
cross-correlation results is to base our correction on a
cross-correlation between all the low redshift ($z\leq 0.6$) objects
that are in the 2QZ catalogue (QSOs, NELGs, LINERs, ...) with the
foreground galaxies. We found 2,667 low-z objects in the NGC of the 2QZ
and the cross-correlation results with $g<21$ galaxies appear in Fig.
15. Note that this result has a much lower amplitude than the $g<21$
galaxy autocorrelation results due to the bigger mis-match between the
average redshift of the $g<21$ and $z<0.6$ 2QZ samples. For the $g<21$
galaxy results this cross-correlation correction used is similarly small
to the auto-correlation corrections discussed above and so we only focus
on the correction for the $r<21$ galaxies of Scranton et al. We again
use the 4\% $z<0.6$ contamination correction appropriate for $g<21$ QSOs
cross-correlated with $r<21$ galaxies. The results are shown in Fig. 16
alongside the uncorrected results and the observational model
($\omega_{qg(\theta)}=-0.024\pm_{0.007}^{0.008}\theta^{-1.0\pm0.3}$)
from Myers et al. (2005) . At $1'$ the correction is -0.0028 and at
$\theta=0'.3$ the correction is -0.004. So at 1$'$ the auto-correlation
function route gives a $2\times$ lower correction than the
cross-correlation route. Since the $r<21$ cross-correlation has poorer
signal to noise while the autocorrelation route has more uncertainty
associated with the assumed $n(z)$ in what follows we shall be using the
average of these two corrections for the QSO-galaxy cross-correlations
of Scranton et al. The (average) corrected results are also shown in
Fig. 16.

\begin{figure}
\begin{center}
\centerline{\epsfxsize = 9.0cm
\epsfbox{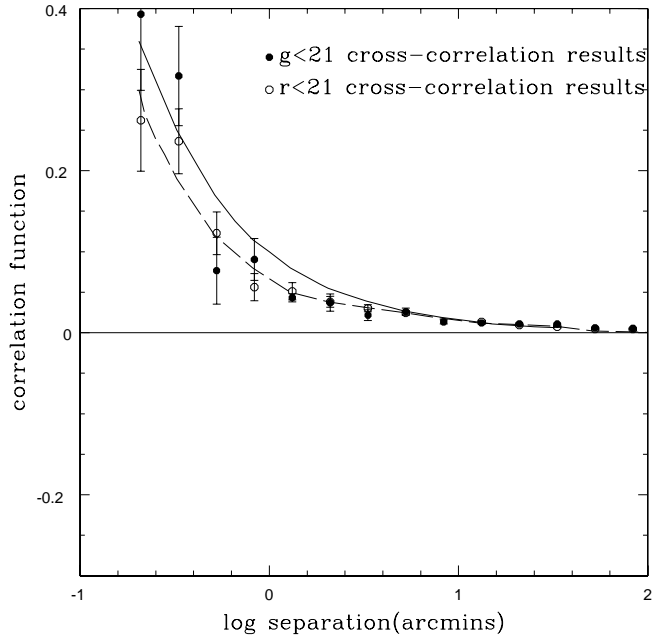}}
\caption{Cross-correlation of the 2,667 low redshift ($z<0.6$) QSOs and
NELGs in the NGC of the 2QZ with our $g<21$ galaxy sample is shown by
the filled circles. The line shows the best fit which is w=0.11$\theta^{-0.8}$. cross-correlation of
the 2,667 low redshift ($z<0.6$) QSOs and NELGs in the NGC of the 2QZ
with our $r<21$ galaxy sample is shown by the open circles. The dashed
line shows the best fit which is w=0.07$\theta^{-0.8}$.}
\label{fig:xi_s2qz}
\end{center}
\end{figure}

\begin{figure}
\begin{center}
\centerline{\epsfxsize = 9.0cm
\epsfbox{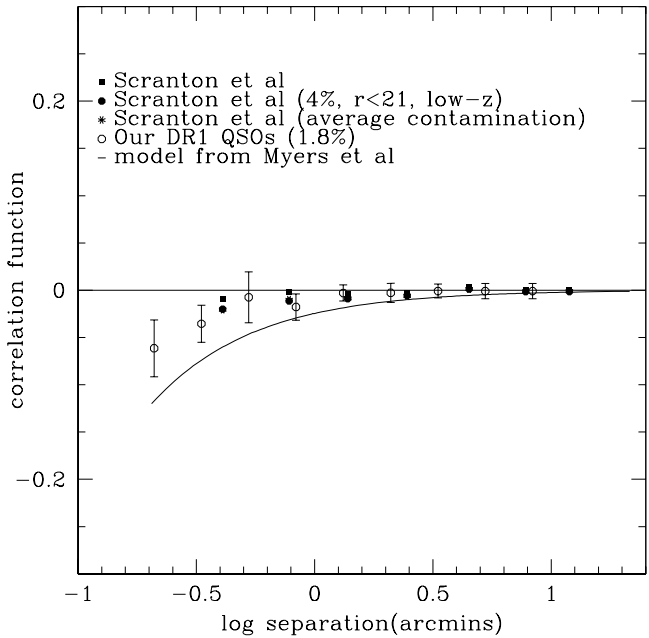}}
\caption{QSO-galaxy cross-correlation results. Squares show original Scranton et al. results, asterisks Scranton et al. assuming
4\% contamination and the fit to the cross-correlation from the low
redshift objects with $r<21$ galaxies. Open circles show our DR1 QSOs
when we apply 4\% contamination and the fit to the cross-correlation from the low redhift objects with $g<21$ galaxies. Filled circles show the results from the average of the two ways of estimating the contamination (galaxy auto-correlation and low-z objects) applied to Scranton et al. Finally, the line
shows the Myers et al.(2005) model for QSO-galaxy cross-correlation.}
\label{fig:xi_s2qz}
\end{center}
\end{figure}



\begin{table}
\caption{CONTAMINATION. DR1 vs. 2QZ}
\centering
\begin{tabular}{lcc}
       \hline
object ID & $1.0\leq z_p\leq  2.2$, $z_s\leq  1.0$ &   $1.0\leq  z_p\leq  2.2$ \\
       \hline \hline
QSOs `11' & 144 & 1999 \\
       \hline
NELGs & 25  & 25   \\
       \hline
Stars '11'& 55 & 55  \\
        \hline
Total number &224& 2079 \\
       \hline \hline \hline
Contamination= &  169/2024= $(8.3\pm0.6)\%$ &\\
        \hline
\end{tabular}
\end{table}

\begin{table}
\caption{Summary of the contamination results}
\centering
\setlength{\tabcolsep}{0.5mm}
\begin{tabular}{lccc}
       \hline
$1.0\leq z_p\leq  2.2$ & $z_s<1.0$ & $z_s<0.6$ &  $z_s<0.3$\\
       \hline \hline
2QZ set & $8.3\pm0.6\%$  & $3.6\pm0.4\%$ & $0.6\pm0.2\%$ \\
       \hline
2QZ+SDSS & $7.7\pm0.2\%$  & $3.8\pm0.2\%$ & $0.9\pm0.1\%$ \\
       \hline
COSMOS & $9.7\pm5.6\%$  & $3.2\pm3.2\%$ & $3.2\pm3.2\%$  \\
       \hline
4 $AA\Omega$ fields& $10.6\pm2.9\%$ & $4.9\pm2.0\%$ & $4.1\pm1.8\%$  \\
       \hline \hline \hline
Total  &$\frac{2}{3}8.3\%+\frac{1}{3}10.6\%$& $\frac{2}{3}3.6\%+\frac{1}{3}4.9\%$&$\frac{2}{3}0.6\%+\frac{1}{3}4.1\%$  \\
cont.= &$9.1\pm0.6\%$&$4.0\pm0.7\%$ &=$1.8\pm0.6\%$  \\
       \hline
\end{tabular}
\end{table}

\section{2QZ versus SDSS comparison - including effect of the galaxy samples}

As noted in Section 2 throughout our analysis we are using $g<21$ 
galaxies whereas Scranton et. al use $r<21$ galaxies. This is the reason
(apart from the contamination of the photometric QSOs) that we don't
expect to get the same answers for the QSO-galaxy cross-correlations. In
Fig. 17 we show our auto-correlation results from $g<21$ galaxies and
the results from $r<21$ galaxies. The galaxy auto-correlation for $g<21$
is taken from Myers et. al (2005); the result for $r<21$ has been
calculated for 35000 SDSS galaxies to this limit. As we see there is a
factor of 2-3 lower anti-correlation for the $r<21$ galaxies. This is
due to increased effects of projection on clustering in the $r<21$
galaxy sample with its 50\% increased depth and 3.5$\times$ higher sky
density (3500deg$^{-2}$ vs. 1000deg$^{-2}$) which leads to weaker
clustering. Therefore we also expect lower lensing anti-correlation
signal by a similar factor of 2-3, based at least on the models used by
Myers et. al (2005) from either Williams \& Irwin (1999) or Gaztanaga
(2004). Using $b\approx0.1$ in a standard $\Lambda$CDM cosmology we
predict the QSO-galaxy cross-correlation result for Myers et al. for
galaxies with $r<21$
($w_{qg}(r<21)=\frac{A_{gg}(r<21)}{A_{gg}(g<21)}w_{qg}(g<21)$ where
$A_{gg}$ is the amplitude of the galaxy auto-correlation function). The
results are shown in Fig. 18. Squares show the Scranton et. al results
(4\% contamination averaged as described in Section 5.7) and the
triangles show Myers et. al results renormalised for an $r<21$ galaxy
sample as described above. As we see the results are in very good
agreement. So the disagreement between the Scranton et al. and Myers et
al results is due to these two factors; contamination was not taken into
account by Scranton et al. and the galaxy samples in the two analyses are
different. When the results are changed based on those two factors they
are consistent.

As already noted, the low signal seen in the corrected cross-correlation
results for bright galaxies could very well be explained by the slope of
the QSO number counts in the region of interest. Table 1 of Scranton et
al (see also Myers et al., 2003) shows the slopes measured for the DR1
sample. At $19<g<19.5$ the slope is $\beta=+0.56$ and at $19.5<g<20.0$
the slope is $\beta=0.43$. From equation (A5) of Myers et al. the
magnification implied for groups by the anti-correlation of
$w_{qg}\simeq-0.15$ at a QSO limit of $g<21$ is $\simeq0.5$ mag for
groups at separation of 1$'$(see Fig. 14). Assuming the former
$\beta=0.56$ slope this would be imply $w_{qc}\simeq0.2$ for the $g<19$
QSO-group correlation in Fig. 9. Assuming the latter $\beta=-0.43$ slope
would imply $w_{qc}=0.04$. This is consistent with the results shown in
Fig. 9 at 1$'$ where the contamination corrected $w_{qc}=0.06\pm0.06$.
The $w_{qg}$ result of Scranton et al. at $17<g<19$ seems to show higher
S/N than any of the others. If the contamination corrections obtained
via Fig. 17 are the same as for the fainter samples, these results will
only be slightly affected, decreasing from $w_{qg}=0.02$ to
$w_{qg}=0.018$ on 1$'$ scales.

The contamination corrected results from Scranton et al. for $17<g<19$
QSOs cross-correlated with $r<21$ galaxies give $w_{qg}=0.014\pm0.009$.
If we take $w_{qg}=-0.01$ for $g<21$ QSOs and the $r<21$ galaxies then
assuming $\beta=0.29$ at $g<21$ and $\beta=0.564$ at $g\simeq19$ this
implies $w=0.015$ from equation (A5) of Myers et al. (2003) and
therefore the bright and faint results are in agreement, confirming the
claim by Scranton et al. even after contamination correction.

In summary, the bright QSO-group galaxy correlation shows positive
signal which is only slightly above the noise but is consistent with the
anticorrelation signal at fainter magnitudes. From here on, we shall
only model the faint QSO cross-correlation where the S/N is generally
higher.

In a further paper we shall cross-correlate the SDSS spectroscopic QSOs
at even brighter limits than discussed here to see if we can detect the
positive correlation expected on the very steep part of the QSO number
counts.

\begin{figure}
\begin{center}
\centerline{\epsfxsize = 9.0cm
\epsfbox{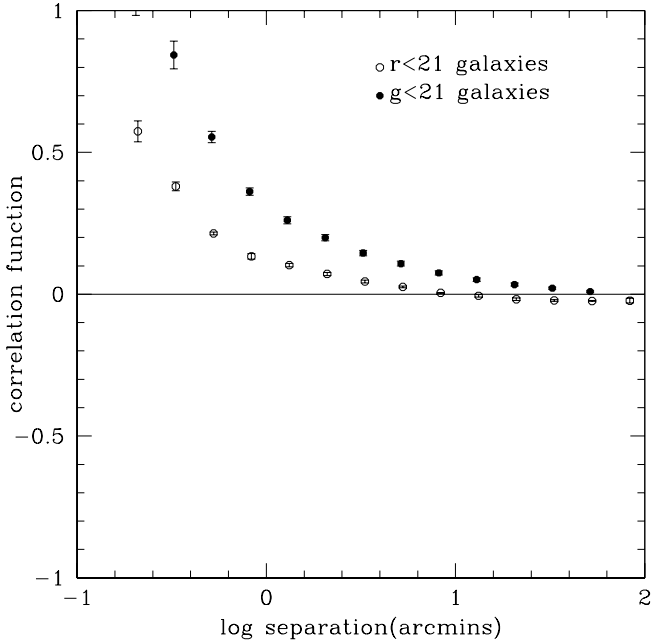}}
\caption{Galaxy auto-correlation results. The filled circles show the
results from the $g<21$ sample taken from Myers et al. (2005). Error bars
represent $1\sigma$ jackknife errors. The open circles show the
auto-correlation results for the $r<21$ sample. The errors are
field-field.}
\label{fig:xi_s2qz}
\end{center}
\end{figure}

\begin{figure}
\begin{center}
\centerline{\epsfxsize = 9.0cm
\epsfbox{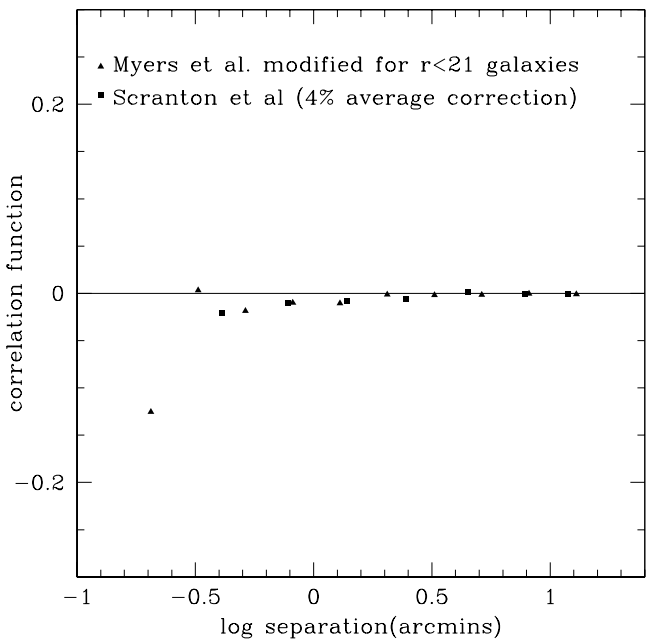}}
\caption{QSO-galaxy cross-correlations. The triangles show Myers et al. (2005) results modified for an $r<21$ galaxy sample. Squares show
the results from Scranton et al. averaged for 4\% contamination as
described in Section 5.7. The two results are in very good agreement.}
\label{fig:xi_s2qz}
\end{center}
\end{figure}

\section{Galaxy model fitting}


Since the observational results appear to be in agreement, the question
then remains as to how the interpretations are so different. Scranton et
al claim that a standard $\Lambda$CM model can explain the QSO lensing
data whereas Myers et al. suggest that $b\approx0.1$ or a high mass
density EdS cosmology (or both) is needed to explain the observations.
Scranton et al. follow the Press-Schechter formalism of Jain, Scranton
and Sheth (2003) assuming the standard cosmology and using the HOD
prescription of Zehavi et al. (2005) as fitted to the SDSS galaxy
correlation function. Previous authors (e.g. Colin et al. 1999) have
suggested that in such models there can be anti-bias at the level of up
to $b\approx0.6$ on scales $0.1<r<1$h$^{-1}$Mpc with $b\approx1$ on
larger scales. As discussed by Myers et al. (2005) significantly stronger
anti-bias than this at $0.1<r<1$h$^{-1}$Mpc was required by their
galaxy-QSO lensing results (see their Fig. 10) and these results are now
strengthened by the agreement found in the new SDSS QSO datasets of
Scranton et al. as well as those analysed here. We believe that the
factor of $\approx6$ that we are seeking is too large to be explained
by some subtlety in the HOD prescription. We also note that Guimaraes,
Myers \& Shanks (2006) used the Hubble Volume simulation of the standard
model to test if some subtlety in either galaxy or group assignment
could explain the large anti-correlation as an artefact. However, these
authors confirmed that approximately unbiased galaxy distributions
assuming the standard cosmology were a factor of $\approx10$ away from
explaining the amplitude of anti-correlation seen with either galaxies
or galaxy groups and clusters.

We take our best data for the spectroscopic 2QZ sample which is the
Northern 2QZ strip reported here, combined with the Southern 2QZ strip
as reported by Myers et al. (2003) for galaxies, inversely weighted by
variance. We find that a fit of
$w_{qg(\theta)}=-0.023\pm{0.006}\theta^{-0.96\pm{0.3}}$ describes our
data, where $\theta$ is expressed in arcminutes. The results are shown
in Fig. 19. Open circles show the weighted average results for both
strips and the short-dashed line shows our model. The new $N+S$ result
is almost exactly the same as that of Myers et al. (see their eq. 21) and
so our fitted galaxy bias remains $b_{0.1}=0.13\pm0.06$ from the model
of Williams \& Irwin (1998). The same model gives $b_{0.1}=0.32\pm0.15$
in the Einstein-de Sitter case. Here and below, the model of Gaztanaga
(2003) would give proportionately smaller bias values (see Table 1 of Myers et
al 2005).

We then take our best photometric DR1 5 areas QSO-galaxy result
corrected for contamination as in Fig. 12. We find that a fit of
$w_{qg(\theta)}=-0.007\pm{0.005}\theta^{-1.4\pm{0.43}}$ describes our
data. In Fig. 19 filled circles show our results and the line shows our
model. Scaling again via equation (19) of Myers et al. we find
$b_{0.1}=0.18\pm0.08$ for the standard cosmology in the Williams and
Irwin case, in reasonable agreement with the result of Myers et al. In
the Einstein-de Sitter case, the result gives $b_{0.1}=0.44\pm0.20$.

Finally, we take the corrected Scranton et al. QSO-galaxy
cross-correlation results (4\% average correction) and find that a fit (Fig. 19) of
$w_{qg(\theta)}=-0.009\pm{0.003}\theta^{-1.0\pm{0.38}}$. The squares in
Fig. 19 show Scranton et al. results with a 4\% average correction and the long dashed line shows our
model. Scaling here via equation (19) of Myers et al. (2005) and taking into
account their $r<21$ galaxy limit gives $b_{0.1}=0.14\pm0.06$ for
the standard cosmology and $b_{0.1}=0.32\pm0.15$ for the Einstein-de
Sitter case under the assumptions of Williams \& Irwin. Again these 
results are in reasonable agreement with those of Myers et al.

\begin{figure}
\begin{center}
\centerline{\epsfxsize = 9.0cm
\epsfbox{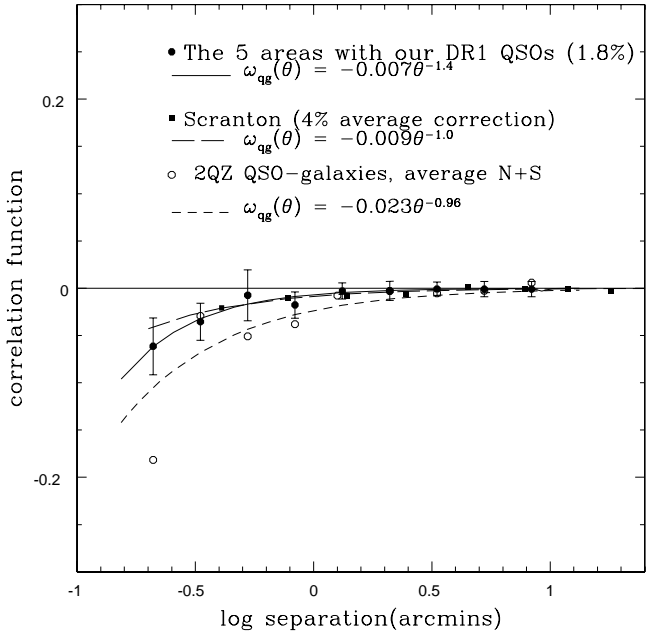}}
\caption{Our fits for our DR1 QSOs in the 5 areas (corrected for 1.8\%
contamination) and for Scranton et al. (4\% average correction based on the $r<21$ galaxy autocorrelation). The weighted averaged 
results from the NGC and SGC (Myers et al.) are shown by the open
circles. The short dashed line shows our fit.}
\label{fig:xi_s2qz}
\end{center}
\end{figure}



\section{Cluster Model Fitting}

Here we fit our best new data in terms of the cluster/group lensing
effect. We, first, follow the modelling procedures of Myers et al. (2003) who
assumed the mass distribution in the groups followed a Singular
Isothermal Sphere (SIS). Here sources are magnified by a factor 

\begin{equation}
\mu =\frac{\theta}{\theta-4\pi\frac{D_{ls}}{D_s}(\frac{\sigma}{c})^2}
\end{equation}

\noindent where $D_{ls}$ is the distance between the lens and the source, $D_s$ is
the distance between the observer and the source, $c$ is the speed of
light, $\sigma$ is the velocity dispersion of the SIS and $\theta$ is
the angle between the source, the observer and the centre of the lens.
The factor $4\pi\frac{D_{ls}}{D_s}(\frac{\sigma}{c})^2$ is the Einstein
radius, so $\theta_E = 4\pi\frac{D_{ls}}{D_s}(\frac{\sigma}{c})^2$. Then
the correlation function is given by $w(\theta) = \mu^{2.5\beta-1}-1$,
where $\beta$ is the slope of the number-magnitude relation. When
$\beta=0.4$ then $w(\theta)=0$, for higher values of $\beta$ we
observe a correlation and for lower values an anti-correlation (Myers et
al). Using the last two equations we can predict the form of the
correlation function.

We also follow Myers et al. (2003) modelling procedures for determing the form of the correlation function based on lensing from dark matter haloes (NFW, Navarro, Frenk \& White 1995). We use the NFW profile which describes haloes that have mass of 9 orders of magnitude, e.g. galaxy clusters. The form of this NFW profile is given (equation A7 in Myers et al. 2003)

\begin{equation}
\rho (r)=\frac{\delta_c\rho _c}{\frac{r}{r_s}(1+\frac{r}{r_s})^2}
\end{equation}

where $\rho _s$ is the critical density and $r_s$ is a representative radial scale. Then, using equations (A2), (A8), (A9), (A10) and (A20) from Myers et al. (2003) we derive the magnification due to the lensing effect that comes from these dark matter haloes.


We take our best photometric DR1 5 areas dataset corrected for
contamination as in Fig. 14 for centres of groups. Using a $\chi^2$ fit
after the fashion of Myers et al., we find that a velocity dispersion of
1040 kms$^{-1}$ fits the $m>7$ group data ($\chi_{red}^2=0.7$) and the SIS
model is shown in Fig. 20 by the solid line. This is close to the value
of 1156kms$^{-1}$ found by Myers et al. We, also fit the NFW profile which is shown by the long dashed line. The best NFW fit has a mass of $M_{1.5}^{DR1}=1.3\times10^{15}h^{-1}M_{\odot}$ with $\chi _{red}^{2}  =1.0$. 

Finally, in a similar way as we did in the previous Section we combine
our results from the cross-correlation of 2QZ QSOs with centres of
groups in the NGC with Myers et al. results in the SGC and then we fit a
model to our weighted average results (open circles, Fig. 20). Here our
$\chi^2$ fit for the SIS velocity dispersion yields 1060 kms$^{-1}$
($\chi_{red}^2=1.3$) again close to the value found by Myers et al. The SIS 
model is shown by the dotted line in Fig. 20. The NFW profile in this case yields a mass of $M_{1.5}^{2QZ}=1.0\times10^{15}h^{-1}M_{\odot}$ with $\chi _{red}^{2}  =1.5$.
 
\begin{figure}
\begin{center}
\centerline{\epsfxsize = 9.0cm
\epsfbox{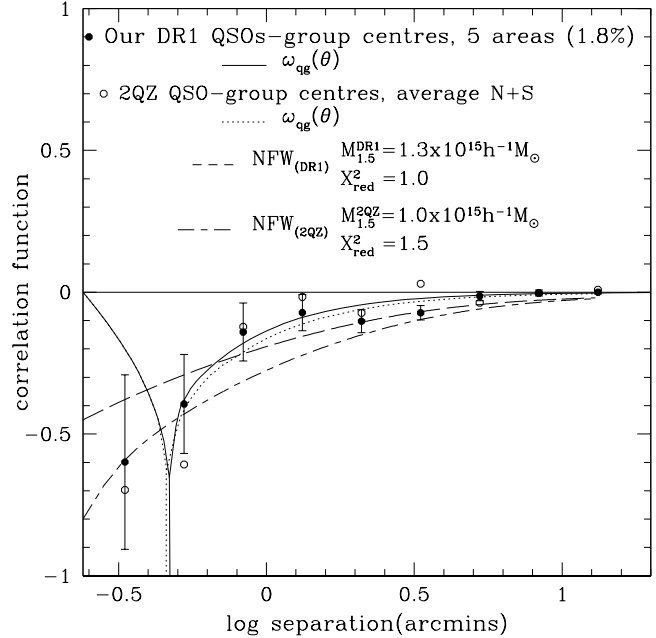}}
\caption{Our weighted average results for both the North and South
strips are shown by the filled circles and the fit to them by the
dotted line. The $\chi^2$ fit for the cross-correlation of our DR1 QSOs
and centres of groups in the 5 areas (corrected for 1.8\% contamination)
is shown by the solid line. 2QZ results are similarly represented by the
open circles and the dotted line. The long dashed line shows the best fitting NFW model profile for the DR1 QSO-group centers cross-correlation and the long-short dashed line for the 2QZ QSO-group centers cross-correlation.}
\label{fig:xi_s2qz}
\end{center}
\end{figure}


\section{Discussion $+$ Conclusion}

Using a photometric QSO sample which was extracted by the same method as
that of Scranton et al. we have investigated the reasons that their
results appear different from Myers et al. (2005), giving a lower
anti-correlation signal.

We first cross-correlated our DR1 SDSS QSO sample with galaxies ($g<21$)
and found that our results are in reasonable agreement with Myers et al.
(2005) at least on scales larger than 1$'$. 2QZ QSOs and DR1 QSOs
cross-correlated with the same $g<21$ galaxy sample give consistent
results on scales larger than 1$'$. The results of Scranton et al.
show a smaller anti-correlation signal as expected from the different
galaxy sample ($r<21$) that they use. In the case of cross-correlations
between QSO-galaxies in clusters, the results repeat the same pattern in
that we detected anti-correlation consistent with the results of Myers
et al. (2003) but at considerably higher S/N than for galaxies.

Then we checked the low-redshift contamination in the $1<z_p<2.2$
photometric SDSS QSO samples. We compared our sample with spectroscopic
samples taken from the 2QZ catalogue, COSMOS field (Prescott et al.) and
AA$\Omega$ spectra that we observed for four new 2dF fields. The
results show $\approx2$\% contamination in our photometric sample for
spectroscopic redshift $z<0.3$ and $\approx4$\% for $z<0.6$. We then
corrected our DR1 results and Scranton et al. results assuming these
percentages of contamination. Then comparing the autocorrelation results
from $g<21$ and $r<21$ galaxies we modified Myers et al. (2005)
QSO-galaxy cross-correlation results and we found that the observational
results are actually in very good agreement.

Therefore we have found that there are two reasons that Scranton et al.
and Myers et al. results look different. The first is that the
low-redshift contamination in the photometric QSO sample has not been taken into
account and the second is that the galaxy sample used in each analysis
is different. {\it When we account for these effects, we consider that
the Scranton et al. SDSS results at faint magnitudes provide strong
observational confirmation of the results of Myers et al. (2005, 2003) in
the same QSO magnitude range.}

Correcting for low-redshift contamination also lowers the positive
correlation claimed by Scranton et al. at bright magnitudes. However, in
the $17<g<19.5$ bin of Scranton et al. some positive signal is still
seen even after contamination correction, at an appropriate amplitude to
match the anti-correlation seen at fainter magnitudes. But the strong
anti-correlation seen at QSO limit $g<21$ suggests that the relevant
slope for $g<19$ QSO samples ($\beta=0.564$) is close to the critical
$\beta=0.4$ slope found at $19<g<20$. This means that little positive
correlation is expected in this magnitude range despite the strong
anti-correlation seen at fainter QSO magnitudes. We note that for
cluster/group lensing the relevant slope may be in the n(m) bin fainter
than the QSO limit bcause this is where the QSOs sit before
magnification.

This then leaves the question of why Myers et al. require a strong
anti-bias of $b\approx0.1$ in the standard cosmology to explain the
faint QSO anti-correlation whereas Scranton et al. require $b\approx1$.
We have noted that Scranton et al. use a more complicated analysis than
that of Myers et al. (2005), using the HOD of Zehavi et al. (2005).
However, the HOD that they use appears to have $b\approx0.6-1$ in the
region of interest and it appears difficult to explain away the factor of
$6-10$ discrepancy by other subtleties in the HOD approach.
Moreover, Guimaraes, Myers and Shanks (2005) used mock galaxy catalogues
with $b\approx1$ derived from the Hubble Volume simulation using
standard parameters and confirmed that models with $b\approx1$ are a
factor of $\approx10$ away from explaining the QSO magnification data.

With the SDSS results now observationally confirming the 2QZ results, we
believe that QSO lensing has now come of age. We note there remain
discrepancies with the $b\approx1$ results found from weak-lensing
cosmic shear results. However as noted by Hirata \& Seljak (2004) there
are serious potential problems with weak lensing shear results in that
the effects of intrinsic galaxy alignments are difficult, if not
impossible, to eliminate from the shear analysis. It is important to
reconcile the QSO magnification and weak shear results for if it proves 
that our QSO magnification results are more accurate then the consequences
for cosmology would be significant. For example, Shanks (2006) has noted 
the potential impact of the QSO magnification results on the interpretation of
the acoustic peaks in the CMB.

\section{Acknowledgments}
We thank A. D. Myers and A.C.C. Guimaraes for useful discussions. 

Funding for the SDSS and SDSS-II has been provided by the Alfred P.
Sloan Foundation, the Participating Institutions, the National Science
Foundation, the U.S. Department of Energy, the National Aeronautics and
Space Administration, the Japanese Monbukagakusho, the Max Planck
Society, and the Higher Education Funding Council for England. The SDSS
Web Site is http://www.sdss.org.

The SDSS is managed by the Astrophysical Research Consortium for the
Participating Institutions. The Participating Institutions are the
American Museum of Natural History, Astrophysical Institute Potsdam,
University of Basel, Cambridge University, Case Western Reserve
University, University of Chicago, Drexel University, Fermilab, the
Institute for Advanced Study, the Japan Participation Group, John
Hopkins University, the Joint Institute for Nuclear Astrophysics, the
Kavli Institute for PArticle Astrophysics and Cosmology, the Korean
Scientist Group, the Chinese Academy of Sciences (LAMOST), Los Alamos
NAtional Observatory, the Max-Planck-Institute for Astronomy (MPIA), the
Max-Planck-Institute for Astrophysics (MPA), New Mexico State
University, Ohio State Univesity, University of Pittburgh, University of
Portsmouth, Princeton University, the United States Naval Observatory,
and tge UNiversity of Washington.

The 2dF QSO Redshift Survey (2QZ) was compiled by the 2QZ survey team
from observations made with the 2-degree Field on the Anglo-Australian
Telescope.

\vspace{10 mm}

\noindent
{\bf References}

\vspace{6 mm}

\noindent
Bartelmann M., 1996, A \& A, 313, 697

\vspace{3 mm}

\noindent
Boyle B. J., Shanks T., Peterson B. A., 1988, MNRAS, 235, 935 

\vspace{3 mm}

\noindent
Broadhurst T., Lehar J., 1995, ApJ, 450, L41

\vspace{3 mm}

\noindent
Colin P., Klypin, A.A., Kravstov, A.V. \& Khokhlov, A.M. 1999, ApJ, 523, 32

\vspace{3 mm}

\noindent
Croom S. M., Shanks T., 1996, MNRAS, 281, 893

\vspace{3 mm}

\noindent
Croom S. M., Shanks T., 1999, MNRAS, 307, L17

\vspace{3 mm}

\noindent
Croom S. M., Smith R. J., Boyle B. J., Shanks T., Miller L., Outram P.J., Loaring N. S., 2004 MNRAS, 349, 1397

\vspace{3 mm}

\noindent
Gaztanaga E, 2003, ApJ, 589, 82

\vspace{3 mm}

\noindent
Guimaraes A.C.C., Myers A.D., Shanks T., 2005, MNRAS, 362, 657

\vspace{3 mm}

\noindent
Hirata C. M., Seljak U., 2004, Phys. Rev. D, 70, 063526

\vspace{3 mm}

\noindent
Jain, B., Scranton, R. \& Sheth, R.K., 2003 MNRAS, 345, 62

\vspace{3 mm}

\noindent
Maoz D., Rix H. W., Gal-Yam A., Gould A., 1997, AJ, 486, 75

\vspace{3 mm}

\noindent
Mellier, Y., Meylan G., 2005, Gravitational Lensing Impact on Cosmology (S225), Edited by Yannick Mellier and Georges Meylan, pp. . ISBN 0521851963. Cambridge, UK: Cambridge University Press, 2005

\vspace{3 mm}

\noindent
Myers A. D., Outram P. J., Shanks T., Boyle B. J., Croom S. M., Loaring N. S., Miller L., Smith R. J., 2003, MNRAS, 342, 467

\vspace{3 mm}

\noindent
Myers A.D., Shanks T., Boyle ÊB.ÊJ., Croom ÊS.ÊM., Loaring ÊN.S., 
Miller ÊL. \& Smith ÊR.J. 2005, MNRAS, 359, 741.

\vspace{3 mm}

\noindent
Myers A. D., 2003, PhD Thesis, University of Durham

\vspace{3 mm}

\noindent 
Myers, A.D. et al., 2007, astro-ph/0612190

\vspace{3 mm}

\noindent
Navarro J. F., Frenk C. S., White S. D. M., 1995, ApJ, 275, 720

\vspace{3 mm}

\noindent
Nollenberg, J. G., Williams L. L. R., 2005, ApJ, 634, 793

\vspace{3 mm} 

\noindent
Peebles P. J. E., 1980, The large scale structure in the Universe, Princeton Univeristy Press, ISBN 0-691-08240-5

\vspace{3 mm}

\noindent
Prescott M. K. M., Impey C. D., Cool R. J., Scoville N.Z., Quasars in the COSMOS field, 2006, ApJ, 644, 100 

\vspace{3 mm}

\noindent
Richards G. T. et al., 2004, ApJS, 155, 257 

\vspace{3 mm}

\noindent
Scranton et al., 2005, ApJ, 633, 589

\vspace{3 mm}

\noindent
Shanks T., 2006, MNRAS submitted, astro-ph/0609339

\vspace{3 mm}

\noindent
Shanks T., Boyle B. J., 1994, MNRAS, 271, 753

\vspace{3 mm}

\noindent
Sharp R.G. et al. 2006, MNRAS submitted, astro-ph/0606137

\vspace{3 mm}

\noindent 
Stevenson P. R. F., Fong R., Shanks T., 1988, MNRAS, 234, 801

\vspace{3 mm}

\noindent
Weinstein M. A. et al., 2004, ApJS, 155, 243

\vspace{3 mm}

\noindent
Williams L. L. R., Irwin M., 1998, MNRAS, 298, 378

\vspace{3 mm}

\noindent
Wu X.P., 1994, A\&A, 286, 748-752

\vspace{3 mm}

\noindent
Zehavi, I. et al., 2005, ApJ, 630, 1

\end{document}